# Blockmodeling of multilevel networks

Aleš Žiberna[*]

*University of Ljubljana*

**Abstract**

The article presents several approaches to the blockmodeling of multilevel network data. Multilevel network data consist of networks that are measured on at least two levels (e.g. between organizations and people) and information on ties between those levels (e.g. information on which people are members of which organizations). Several approaches will be considered: a separate analysis of the levels; transforming all networks to one level and blockmodeling on this level using information from all levels; and a truly multilevel approach where all levels and ties among them are modeled at the same time. Advantages and disadvantages of these approaches will be discussed.

**Keywords: multilevel networks, multilevel analysis, generalised blockmodeling**

## 1    Introduction

In the article several approaches to blockmodeling multilevel networks are presented. First, the type of data that are referred to here as multilevel networks will be introduced, followed by an explanation of the problem that the blockmodeling of multilevel networks ("multilevel blockmodeling") should solve. Multilevel networks are composed of one-mode networks (possibly multi-relational) for each mode and two-mode networks that "join" units from different levels.

---

[*] Aleš Žiberna, Faculty of Social Sciences, University of Ljubljana, Kardeljeva ploščad 5, 1000 Ljubljana, Slovenia; E-mail: ales.ziberna@fdv.uni-lj.si; ales.ziberna@gmail.comi

The goal of multilevel blockmodeling is then to find a blockmodel (groups and ties among them) for all these networks simultaneously, namely to partition the units at all levels into groups by taking all available information into account and determining the ties among these groups.

Three general approaches are presented: a separate analysis of each mode and a comparison of results; conversion of the multilevel problem to a classical one-level blockmodeling problem; and a true multilevel approach. The suggested approaches are applied to a specific example. At the end of the article, the advantages and disadvantages of the suggested approaches are discussed.

## 2 Multilevel networks

Multilevel networks can be defined in several different ways. The document "What Are Multilevel Networks" prepared by the Multi-level Network Modeling Group (MNMG) (2012) identifies four different definitions of multilevel networks or multilevel approaches to the analysis of networks. In this article, the fourth definition is used, namely the one where ties between units at each level are studied together with ties between levels. Therefore, multilevel networks are defined here as networks composed of one-mode networks (possibly multi-relational) for each mode and two-mode networks that "join" units from different levels. While most (if not all) multilevel networks are also multi-relational, these two concepts should not be confused. Units in a multilevel network are composed of different types of units, where each type corresponds to a *level*. On the other hand, a multi-relational network is a network where several *relations* are measured on one (or more) sets of units.

Let me first introduce some notation:

- $U = \{u_1, u_2, \ldots, u_n\}$ is a set of all units and *n* is the number of all units.
- *U* can be partitioned into *L* distinct sets (usually two) that represent different levels, so that $U = \bigcup_{i=1}^{L} U_i$ and $U_i \cap U_j = \emptyset$, for each $i \neq j$.
- Each level has $n_l = |U_l|$ units, $l \in [1, L]$
- *R* is a set of *K* relations, where relations are denoted by $R^k, k \in [1, K]$ and are measured on units *U*.
- Relations can be defined on all levels (on units from all sets) or on only some levels or combinations of levels.
- A multilevel multi-relational network is denoted by *N* = (*U, R*).
- One-mode networks are denoted by $N_l = (U_l, R')$, where $l \in [1, L]$ and $R' \subseteq R$. If only certain relations are used, this is indicated by superscript.
- Two-mode networks are denoted by $N_{ij} = (U_i, U_j, R')$, where $i \in [1, L]$, $j \in [1, L]$, $i \neq j$ and $R' \subseteq R$. If only certain relations are used, this is indicated by superscript.
- The relation $R^k$ is represented by a valued matrix $\mathbf{R^k}$ with elements $[r_{ij}^k]$, where value $r_{ij}^k$ indicates the value (or weight) for the arc from unit *i* to unit *j* on relation *k*.

- Each matrix $\mathbf{R}^k, k \in [1, K]$ can be partitioned into matrices $\mathbf{R}^k_{ij}$, where the first and second dimensions correspond to sets of units/levels. If a relation $R^k$ is not defined from set $U_i$ to set $U_j$ then all entries of submatrix $\mathbf{R}^k_{ij}$ are undefined[1].
- Several relations are represented by three-way arrays $\mathbf{R}^{k_1, k_2, \ldots}$ (this is essentially the same as the multiway matrices from Borgatti and Everett, 1992), where the included relations ($k_1$ and $k_2$, …) are listed in superscript. If all relations between relations $k_1$ and $k_2$ are included, this can be written as $\mathbf{R}^{k_1 - k_2}$. Such arrays can be partitioned according to levels in the same way as matrices representing single matrices.
- The whole multilevel network *N* can be represented by a three-way array $\mathbf{R} = \mathbf{R}^{1-K}$.
- The relations can be binary, valued, signed or any other kind for which an appropriate blockmodeling approach is defined. The blockmodeling approach used to blockmodel a certain relation must be applicable to such a relation.

While all the methods suggested here can in theory[2] be used on any number of levels, I limit myself to two levels in the whole example section and some other parts of this article. In those parts, I always explicitly state that I am discussing the two-level case. Most of the discussion in the article is limited to the case where two-mode networks represent partitions of "lower" level units into "higher" level units, that is each "lower" level unit is tied to exactly one "higher" level unit, while each "higher" level unit is tied to at least one "lower" level unit. The methodology suggested here is suitable for all types of two-mode networks, although especially the discussions on reshaping networks and modeling two-mode networks are largely conditional on this assumption as the most likely case in a multilevel context.

When I limit the discussion to two-level networks, I also limit myself to the case with only three relations. Therefore, I restrict the discussion to the case of a two-level, three-relational network, where relation $R_1$ is defined on set $U_1$ (a set of individuals), relation $R_2$ is on set $U_2$ (a set of institutions), and relation $R_3$ from set $U_1$ to set $U_2$. The multilevel (and multi-relational) network *N* is represented by matrix **R** that can be split into the following defined[3] submatrices:

- submatrix $\mathbf{R}^1_{11}$, representing the one-mode network of individuals/first-level units $N_1 = (U_1, R^1)$
- submatrix $\mathbf{R}^2_{22}$ representing the one-mode network of institutions/second-level units $N_2 = (U_2, R^2)$
- submatrix $\mathbf{R}^3_{12}$, a two-mode (affiliation) network tying individuals to institutions / ties between first- and second-level units $T = N_{12} = (U_1, U_2, R^3)$

---

[1] In our implementation they are coded as 0, although this is irrelevant since they are ignored in all computations.

[2] In practice, the time complexity of the algorithm and the complexity of the numerous "interactions" between levels would prohibit the application of the method to many levels (e.g., more than 3 or 4).

[3] Matrices whose entries are defined. The remaining matrices $(\mathbf{R}^1_{22}, \mathbf{R}^1_{21}, \mathbf{R}^1_{12}, \mathbf{R}^2_{11}, \mathbf{R}^2_{12}, \mathbf{R}^2_{21}, \mathbf{R}^3_{11}, \mathbf{R}^3_{11}, \mathbf{R}^3_{21})$ are undefined (in practice coded as 0 matrices).

To the best of my knowledge, Iacobucci and Wasserman (1990) were the first to suggest the analysis of such networks and they soon (Wasserman and Iacobucci, 1991) also presented a method for the statistical modeling of such networks. The importance of a multilevel view was later advocated by Brass et al. (2004). However, I am aware of only one example of a multilevel network dataset, the one gathered and analyzed by Lazega et al. (2006, 2008, 2013). Recently, exponential random effects models were also extended to multilevel networks (Wang et al., 2013), where they used the same dataset to demonstrate the importance of the method. The same dataset is also used here in the example section.

Additional methods (e.g. Snijders et al., 2013) and applications (e.g. Bellotti, 2012; Snijders et al., 2013) can be found for combinations of only one-mode networks at one level and a two-mode network connecting this level to another level. Such networks can be seen as a special case of multilevel networks as defined here where no relations are collected for one level.

## 3   Blockmodeling and some of its extensions

Blockmodeling aims to partition network units into clusters and, at the same time, to partition the set of ties into blocks (Doreian et al., 2005a, p. 29). Blockmodeling can be also be "[v]iewed as a method of data reduction, […] a valuable technique in which redundant elements in an observed system are reduced to yield a simplified model of relationships among types of elements (units)" (Borgatti and Everett, 1992). There are several approaches to blockmodeling, such as stochastic blockmodeling (Holland et al., 1983; Anderson et al., 1992; Snijders and Nowicki, 1997), conventional blockmodeling (e.g. Breiger et al., 1975; Burt, 1976; see Doreian et al., 2005a, pp. 25–26 for definition) and generalized blockmodeling (Doreian et al., 2005a, 1994). While this article focuses on generalized blockmodeling (Doreian et al., 2005a), more precisely homogeneity blockmodeling (Žiberna, 2007), at least the first two suggested approaches (separate analysis and conversion to one-level blockmodeling) can be easily implemented using other approaches.

Some additional notation is introduced here:

- $C_i$ is a cluster of units for $1 \leq i \leq m$, where $m$ is the number of clusters.
- $\mathbf{C} = \{C_1, C_2, \ldots, C_m\}$ is a partition of the set $U$; $\bigcup_{i=1}^{m} C_i = U$; $C_i \cap C_j = 0, i \neq j$.
- $P(\mathbf{C}, E, N)$ is the value of a criterion function that measures the fit of partition **C** and equivalence $E$ to network $N$. $E$ can be expressed in different terms, e.g. allowed block types, a pre-specified image etc. (see Doreian et al., 1994 for a further discussion) but it must also include the type of blockmodeling (e.g. binary, sum of squares etc. – see Žiberna, 2007 for a further discussion).

In generalized blockmodeling the criterion function in optimized when searching for the optimal **C** given the $E$ and $N$. A computation of the criterion function for single-relational networks is described in works presenting different approaches to generalized blockmodeling (e.g. Doreian et al., 2005a; Žiberna, 2007).

In the remainder of this section, I present several extensions to generalized blockmodeling that are required for the approaches suggested in the next section, especially for the true multilevel approach, although their usefulness extends well beyond their application to multilevel blockmodeling.

## 3.1 Multi-relational blockmodeling

Although (classical) blockmodeling was initially developed for multi-relational networks (Breiger et al., 1975; Burt, 1976; White and Reitz, 1983; White et al., 1976), generalized blockmodeling was only developed for single relations. While Doreian et al. (Doreian et al., 2005a; Ferligoj et al., 1996) discussed multiple relations among possible extensions to generalized blockmodeling in their book (Doreian et al., 2005a), they did so with serious reservations. Recently, Brusco et al. (2013) presented multi-objective blockmodeling that can be used to blockmodel multi-relational networks. Even though their approach is most likely more appropriate, a simpler approach is used here.

The extension of generalized blockmodeling to multiple relations is at least in technical terms straightforward. Generalized blockmodeling is an optimization approach that searches for the optimal partition by minimizing the criterion function. We could say that in the case of multiple relations this turns into a multi-objective clustering problem (Ferligoj and Batagelj, 1992) (a criterion function for each relation presenting one objective). One possibility that will be used in this article is to transform this multi-objective problem to single-objective problem using the weighted sum approach (Ehrgott and Wiecek, 2005; Ferligoj and Batagelj, 1992). Several issues arise when using this approach, from choosing suitable weights to purely conceptual problems; however, these issues exceed the scope of this article.

The criterion function for multi-relational network *N* with *K* relations can be computed as follows:

$$P(\boldsymbol{C}, E, N) = \sum_{k=1}^{K} w_k P(\boldsymbol{C}, E^k, N^k), \tag{1}$$

where $w_k \geq 0$ is a weight for relation $R^k, k \in [1, K]$

As mentioned, the multi-objective approach (Brusco et al., 2013) might be more appropriate than the weighted sum approach, but I have currently not yet implemented it in my software.

## 3.2 Different sets of units

In multilevel blockmodeling we have several sets of units (at least two), one for each level. These sets of units must be partitioned separately, that is units from different sets cannot be together in the same cluster.

Let us define the set of feasible partitions $\Phi$ as:

$\mathbf{C} \in \Phi$ iff:

$$\mathbf{C} = \bigcup_{l=1}^{L} \mathbf{C}_l \tag{2}$$

$\forall \mathbf{C}_l \subset \mathbf{C}, \mathbf{C}_l = \{C_{l,1}, \ldots, C_{l,m_l}\}, \bigcup_{i=1}^{m_l} C_{l,i} = U_l$, where $\mathbf{C}_l$ is a partition of the set $U_l$, $C_{l,i}$ is the $i$-th cluster of the partition $\mathbf{C}_l$ and $m_l$ is the number of clusters in the partition $\mathbf{C}_l$.

$U_i \cap U_j = \emptyset, \forall\, i \neq j$

The problem can be expressed as a constrained clustering problem (Batagelj and Ferligoj, 1998; Gordon, 1996).

$$P(\mathbf{C}^*, E, N) = \min_{\mathbf{C} \in \Phi} P(\mathbf{C}, E, N) \tag{3}$$

Such a restriction for two sets is already used in two-mode blockmodeling (e.g. Doreian et al., 2004). For multilevel blockmodeling such a restriction must be extended to single-mode networks and more than two sets. The usability of such restrictions goes beyond multilevel blockmodeling. It can be used always when distinct sets of units exist that either should not be mixed or we believe that the optimal partition will not have them mixed. When this restriction is used, this reduces the neighborhood that must be searched in either a local search or similar algorithm, thus reducing its time complexity. For example, such a restriction could be used when analyzing a baboon grooming network as was done by Doreian et al. (2005a, 2005b) since baboons of different genders never appear in the same cluster.

## 4 Multilevel blockmodeling

The ultimate goal of multilevel blockmodeling is to find a blockmodel (groups and ties among them) for the whole multilevel network, which is to partition the units at all levels into groups by taking all available information into account and determining the ties among these groups.

In this article, three general approaches are discussed:

a) a separate analysis of each mode and a comparison of the results;
b) conversion of the multilevel problem to a classical one-level blockmodeling problem (hereafter "the conversion approach"); and
c) a true multilevel approach.

These are not really alternative approaches since at least the first one (separate analysis) should be the first step in any blockmodeling analysis of multilevel networks. The separate analysis approach (a) represents a good exploratory technique and can guide a more complex analysis and show whether more complex approaches are even justified. The conversions approach (b) is appropriate when we want to focus on a certain level while using information from the other level(s) to improve the partition and/or when the other level(s) can be used as indirect relations for units of the level in focus. In contrast, the multilevel approach (c) should be used when we already have some knowledge about the network's structure. It can provide us with a novel insight into the ties among clusters from different levels. It can also help us search for such clusters at individual levels in such a way that the ties among them are relatively "clean". In addition, the multilevel approach can have similar effects as the conversion approach since information from one level is used to better determine clusters on the other level.

Use of the first and at least one of the other two approaches is also in line with the idea of Lazega et al. (2013) that it "is important to examine both levels separately and jointly".

## 4.1 A separate analysis of each mode and a comparison of the results

The simplest way to analyze a multilevel network using blockmodeling is to blockmodel each level separately and then compare the results. The comparison can be done in several ways:

a) forcing the partition obtained at one level onto the other level(s) and analyzing the fit; or
b) obtaining the partitions on all levels and comparing them.

Both options are complementary and preferably both should be used. The first option in (a) means that, after obtaining a partition on a given level, this partition is forced onto another level. This can be done by either reshaping the partition to the level on which it is to be forced or reshaping the one-mode network of the level on which the partition is to be forced to the level on which the partition was obtained. Both reshapings are done through the use of the two-mode networks joining the two levels.

The more detailed description that follows applies to the case of a two-level, three-relational network. The reshaping is most straightforward when the two-mode network essentially represents a partition of units of the first level into classes defined by the second level and we are reshaping the second-level partition to the first level. In such cases, the second-level partition can be reshaped to the first level simply by assigning to the units of the first level the class (cluster) of the units of the second level to which these units belong.

Similarly, we can easily reshape the network of the second level to the first by assigning the tie of the second-level units to pairs of first-level units that are associated with these second-level units. This can be simply obtained by pre- and post-multiplying the matrix representing the second-level network by the matrix representing the two-mode network (transposed when needed) as presented in Equation ( 4 ).

$$\mathbf{R_{11}^{2^*}} = \mathbf{R_{12}^{3}} \times \mathbf{R_{22}^{2}} \times \left(\mathbf{R_{12}^{3}}\right)', \text{where } ' \text{ is the transpose operation.} \tag{4}$$

The reshaped network $N_1^{2^*} = (U_1, R^{2^*})$ actually represents indirect ties between units of the first level through the ties among second-level units to which these first-level units are associated. Such a transformation is also undertaken by Lazega et al. (2013) where they call neighbors in such a resulting network "dual actors".

The transformations are a little more complicated in the other direction or when first-level units are tied to more than one second-level unit. In such cases, some averaging, voting or aggregation rules are required[4].

After a partition at one level is obtained and a suitable reshaping has been applied, we can see how this partition fits the other level. That is, we can check whether the pattern of ties of the second network is well explained by this partition and therefore by the structure of the first-level network. We could say that we are performing a kind of pre-specified blockmodeling (Batagelj et al., 1998) and checking the fit of the pre-specified partition (and possibly a blockmodel image) to a network. If the fit is good (significantly better than random), we can say that the structures of both networks are associated. In addition, we can check whether the blockmodel images are similar at both levels. If they are, this indicates that not only are the groups on one level associated with the groups on the other level, but so too is the pattern of ties among groups.

The second option (b) is to compare partitions obtained at both levels. This is done by reshaping one of the partitions for it to be compatible with the other and using some classical indices for comparing partitions to compare them, such as the Rand Index (Rand, 1971) or Adjusted Rand Index (ARI) (Hubert and Arabie, 1985). Obviously, larger values of these indices indicate a stronger association among the partitions and therefore among the global structures of the one-mode networks at different levels. All values of ARI over 0 indicate that the association is greater than would be expected by chance.

Since this approach is a good exploratory technique, it is simple to perform and allows an estimation of the association of group structures across levels, it should always be the first step in the analysis. These comparisons allow us to determine whether there is some similarity in the structure of the two networks and whether the similarity is only in the partitions or also in the pattern of ties among groups. Where no similarity is found, more complex analyses are probably not justified. In case of only partition similarities, the single-relational version (explained in the next subsection) of the conversion approach is most likely unsuitable.

Of course, this approach also has limitations especially since all partitions are only based on one level and that the ties between groups of different levels cannot be modeled, only observed. However, this does not limit its usefulness as an exploratory technique.

---

[4] For example, if a first-level unit belongs to several second-level units and we want to reshape the second-level partition to the first level, there is a problem of what class to assign to this first-level unit if all the second-level units have different classes. One possibility is to assign a majority class if such a class exists, to randomly select one class, or to create a new class for each unique combination of classes of the affiliated second-level units.

Similarly, the reshaping of the network in such a case requires some aggregation principles to determine the presence or value of ties in the reshaped network. For valued networks, sum, average, minimum or maximum are possible aggregating functions, whereas when a binary network is used some threshold could be supplied to determine at what density of ties in the subnetwork of second-level units the tie would be formed in the reshaped network. As this exceeds the scope of this paper, any more detailed discussion is omitted.

## 4.2 Conversion of the multilevel problem to a classical one-level blockmodeling problem

The first approach suggested here that takes information about all levels into account is to convert this multilevel problem to a one-level problem. The approach is appropriate in cases where we believe that the partitions at different levels are practically the same (after reshaping) and we want to use as much information as possible to find these partitions. In fact, when using this approach only a partition at one level is obtained (the "main" level), which can then be reshaped if desired to obtain partitions at "other"[5] levels[6]. Therefore, we should only use it if we find in the separate analysis stage that the partitions for all levels are similar or if that one partition at least approximately fits all levels.

In this approach, we therefore reshape the network from "other" levels to the "main" level and then partition all networks at the main level simultaneously. The reshaped networks represent additional (indirect) relations[7] in the "main" level's network. If, in the case of two levels, we reshape $N_2^2$ into $N_1^{2*}$ as was presented in the previous subsection, the "joint" multi-relational network is $N_1^{1,2*} = N_1^1 \cup N_1^{2*} = (U_1, R^1, R^{2*})$.

We have two options when analyzing the obtained multi-relational network. The first one is to somehow aggregate these relations by using some function like maximum (other options include minimum, average and sum) on relations on the same tie. This option only really makes sense if all networks measure similar concepts and have a similar structure in terms of both the partitions and patterns of ties among groups.

For example, if we can consider one network person's direct access to some resources and the other network person's indirect access through institutional exchange. In some cases, it might be sensible to find a partition at one level using this approach, but not on the other. E.g. it might make sense to assume that employees can access resources through their firm's connections, but not vice versa[8]. In this case, when partitioning the employees it would be sensible to include their firms' connections to better estimate their position in some network, but it would not make sense to estimate the firms' positions also using their employees' connections.

---

[5] Other than the "main" level, that is other than the one to which all networks were reshaped.

[6] In the theoretical part, if not explicitly stated I otherwise discuss the more general case where there can be two or more levels. This means that there can be one or more »other« levels and therefore either singular or plural form are appropriate. I will however use the plural form with the understanding that in the case of just two levels there is only one "other" level.

[7] In the case where the network to be reshaped is multi-relational, we also obtain several relations by reshaping each relation separately.

[8] I do not imply that firms can (never) access resources through employees' (personal) networks.

The second option, which is usually more appropriate, is to blockmodel the multi-relational network directly using the multi-relational blockmodeling discussed in subsection 3.1. This simply means that we perform blockmodeling on all relations simultaneously by constraining them to the same partition using Equation ( 1 ).

In the two-level case, the criterion function used is then:

$$P(\mathbf{C}, E, N_1^{1,2^*}) = w_1 P(\mathbf{C}, E^1, N_1^1) + w_2 P(\mathbf{C}, E^{2^*}, N_1^{2^*}) \tag{5}$$

The advantages of this approach are that it is still relatively simple to perform and that it uses all available information (on all) levels to obtain a partition at the selected ("main") level. However, as discussed above, this only makes sense in certain cases. The approach also has several disadvantages, the first being that some information is lost in the aggregation, especially if the single-relational approach (aggregating relations prior to the blockmodeling analysis) is used. Second, the choice of suitable weights can be problematic when a multi-relational version is used. Finally, the approach obtains just one[9] partition that is then reshaped to different levels. This means that the ties between groups at different levels are fixed and cannot be observed or modeled. The "other" levels' partitions are a function of the original partition obtained at the "main" level and the two-mode network(s) joining the "other" levels with the "main" level.

## 4.3  The true multilevel approach

The purpose of this approach is to partition units of all levels simultaneously (using multi-criteria clustering) by taking account of both the ties within levels and those between levels (two-mode network(s)). Formally for the two-level, three-relational case (network $N_1$, network $N_2$ and network $N_{12}$) this means finding partitions $\mathbf{C_1}$ (of set $U_1$) and $\mathbf{C_2}$ (of set $U_2$) that optimize the following criterion function:

$$\begin{aligned} P(\mathbf{C_1}, \mathbf{C_2}, E^1, E^2, E^{12}, N_1, N_2, N_3) \\ = w_1 P(\mathbf{C_1}, E^1, N_1) + w_2 P(\mathbf{C_2}, E^2, N_2) + w_3 P(\mathbf{C_1}, \mathbf{C_2}, E^{12}, N_{12}) \end{aligned} \tag{6}$$

A more general approach (not adapted to a certain number of levels or relations) is to join all one-mode and two-mode networks into a single multi-relational network[10] $N$ (also represented by three-way array **R**) as introduced in Section 2. In this case, the criterion function is simply the criterion function for multi-relational blockmodeling (Equation ( 1 )) with the constraint that each level is partitioned separately (Equation ( 3 )).

---

[9] We could use different levels as a »base« level, that is the level to which other levels are reshaped. The partitions obtained using different base levels might then slightly differ when reshaped to the same level, especially if in the two-mode network units from both sets of nodes can have many ties (to nodes of the other set).

[10] Although in some applications it might be possible to treat the whole multilevel network as a one-relational network, this is relatively unlikely as most likely one-mode networks on different levels and the two-mode network(s) will measure different relations in the majority of applications. Therefore, the more general and probable situation where the multilevel network is also multi-relational will be considered here.

### 4.3.1 Specifying equivalences for parts of the multilevel network

As can be seen from Equation ( 1 ), we need to specify equivalence for each relation. Each relation is usually defined only on one set/level of units (single-level one-mode network) or only between two sets of units (two-mode networks). For parts of the network where the relations are not defined, the only allowed block type should be a "Do not care" block (Doreian et al., 2005a, p. 235). The inconsistency of such a block is always zero (regardless of the ties in the block). This ensures that only appropriate parts of the relations/network or in technical terms of the three-way array **R** are taken into account when computing the value of the criterion function.

For these parts of relations that are defined, that is for each of the one-mode and two-mode networks separately, we can specify suitable generalized blockmodeling approaches, allowed block types and possibly their positions (or equivalences). Suitable specifications for one-mode networks can be found in the relevant literature (e.g. Doreian et al., 2005a; Žiberna, 2007). While generalized blockmodeling of two-mode networks has also been covered (Doreian et al., 2005a, 2004), some aspects specific to its use in multilevel blockmodeling are discussed here.

In most cases it will be desired that most blocks in the two-mode network are null (empty, without any ties) since this makes the connections between the groups at different levels clearer. Preferably, such blocks would have no inconsistencies (no ties). The way of obtaining such null blocks depends on the blockmodeling approach applied. Using some approaches (like those shown in the example in Section 5.4), almost perfect null blocks in two-mode networks can be obtained by using structural equivalence and simply giving a large weight to the two-mode network criterion function. The goal of very few or even no inconsistencies in null blocks can be achieved by heavily penalizing the inconsistencies in the null blocks as e.g. in Doreian et al. (2005a, pp. 260–261).  However, sometimes we might also want to impose restrictions in terms of pre-specified blockmodels (Batagelj et al., 1998) or allowed images, e.g. that each row (i.e. level one) cluster might be associated with only one column (i.e. level two) cluster. A further discussion of possible restrictions can be found in Appendix A.  In the case of two levels when equivalences for two-mode networks ensure (at least approximately) that each lower level cluster is tied (actually the units it contains) to only one higher level cluster and vice versa, this restriction is similar to the restriction imposed by the multi-relational conversion approach (see Appendix A for details) and we therefore expect these approaches to produce similar results (of course, provided that equivalences for the one-mode networks are also specified in the same way for the two approaches).

### 4.3.2 Advantages and disadvantages of the true multilevel approach

The true multilevel approach has several advantages, namely that it takes all available information (all one-mode and two-mode networks) into account, that no aggregation is necessary, and that ties between levels can be modeled. However, it also has several drawbacks. In conceptual terms, the main disadvantages are that there are no clear guidelines concerning what are appropriate restrictions for ties between levels and what are appropriate weights for different parts of multi-relational networks, that is for level-specific one-mode networks and for the two-mode networks. In the event of equal weights, in principle the parts with larger inconsistencies have a bigger influence on the results. As the inconsistencies are dependent on the equivalences (or allowed block types and their positions), networks' size, pattern of and number of clusters, all these factors influence the appropriate weighting. A suggestion that is also followed in the example in subsection 5.4 is to make the weights reciprocally proportional to the inconsistencies of the relations/single-relational network if the whole network (for which the relation is defined) would be in a single block (that is, if all units (from the same mode/level in two-mode networks) would be in the same cluster). If we want very few inconsistencies in the two-mode network(s), this/these network(s) can be given higher weight(s) (e.g. double the one computed based on the suggestion in this paragraph as is also used in the example).

Additional disadvantages are tied to optimizational problems, especially as a local search with a single exchange and move as allowed "moves" is currently used for optimization. Finding an optimal partition using the direct approach is in most cases an NP-hard problem (Batagelj et al., 2004, p. 461). The multilevel approach is even more time-demanding as there are more units in a multilevel network than in single-level networks. However, the main problem lies in the fact that currently a local search with allowed transformations being a single exchange and a single move is used (see e.g. Batagelj et al., 1992, p. 127 for details). This is problematic since in the multilevel approach quite hard constraints are usually desired for a two-mode network(s), typically by desiring null blocks and strongly weighting the inconsistencies in the two-mode network(s) (at least those in the null blocks)[11]. If the current partition is such that ties between a certain higher level unit and some lower level units are in a non-null block, moving just the higher level unit (since only one move at a time is allowed) would most likely move several ties in the two-mode network to the null block and would therefore be very costly[12], to a such an extent that the move would most likely not be selected. In the current implementation, I attempt to circumvent this problem by applying two strategies. The first one is brute force, namely by using many random starting partitions with a local search. The second one is not to weight inconsistencies in the two-mode network(s) too strongly[13] in the first stage in order not to make such moves too costly. If this results in an under-structured two-mode network(s) (too many ties in the null block or too few null blocks), the resulting partition can be further optimized with more stringent constraints on the two-mode network(s), namely by increasing the weight of the inconsistencies in the two-mode network(s). Of course, it would be better to use an adapted tabu search (Brusco and Steinley, 2011) or similar algorithm that would temporarily allow costly moves or direct multiobjective blockmodeling (Brusco et al., 2013).

---

[11] Meaning that an additional tie in the null block of the two-mode network increases the inconsistency by much more than the inconsistency in the one-mode networks

[12] Meaning that they do not increase the inconsistency too much

[13] E.g. to use weights computed as suggested in the previous paragraph

# 5   Example: Application to a multilevel network of elite cancer researchers in France

The suggested approach is demonstrated on a multilevel network of the elite of cancer researchers in France (Lazega et al., 2008). The analyzed multilevel network is composed of two levels, a level of researchers and a level of research labs. The networks and other data used are described in more detail in the following subsection. Generalized blockmodeling offers a wide range of possible analysis. Due to space limitations and the focus on the method (not the application) of this article, only one possibility is presented here. An attempt is made to find cohesive groups and determine whether they are associated with certain researchers' or labs' specialties.

To achieve this, generalized blockmodeling with pre-specified blockmodeling was used. The pre-specified blockmodel corresponding to cohesive groups was used for one-mode networks, namely by only allowing "null" blocks in off-diagonal blocks and only "complete" blocks on the diagonal blocks within each level and/or relation. Several approaches to generalized blockmodeling exist (Doreian et al., 2005a; Žiberna, 2007). Homogeneity blockmodeling with sum of squares (SS) blockmodeling according to structural equivalence was used here in all the analyses for all levels and relations. However, it should be emphasized that there is no need to use the same approach for all levels/relations. In order to prevent null blocks from appearing where complete blocks are required (on the diagonal) and to prevent almost null blocks from being classified as complete, the constrained version of complete blocks as described in Žiberna(2013a) was used. That is, the value from which sum of square deviations were computed was constrained to a pre-specified value, which was set to twice the mean of the relation/level. As a local search is used to find the "optimal" partition, at least 1,000 random starting points were used in all analyses. All of the analysis was performed using the development version of `blockmodeling` 0.2.2 package (Žiberna, 2013b, 2013c) within the R 3.0.1 software environment for statistical computing and graphics (R Core Team, 2013).

## 5.1   Data description

The suggested approaches were applied to the multilevel network of the elite of cancer researchers in France gathered and analyzed by Lazega et al. (Lazega et al., 2013, 2008). Several networks of researchers and several networks of labs were collected together with a two-mode network of researchers' membership in laboratories (labs). For this demonstration, the same kind of aggregation as performed by Lazega et al. (2008) was used.

This gave us the following networks:

- o   a network of researchers
- o   a network of labs
- o   a two-mode network of labs and researchers: A membership matrix of labs x researchers

In this application I am using data on 78 labs and 98 researchers, namely all cases where I had data on pairs of researchers and labs (or larger groups since there can be more than one researcher per lab). While some labs and researchers have no outgoing ties, they were not excluded since they were nominated by others.

First, both networks (and the ties between them) are presented graphically in Figure 1 and Figure 2. Little can be learned from these two representations except that the network of researchers is denser and perhaps has more structure. Table 1 reveals important differences between the networks. The reciprocity and to a lesser extent clustering coefficient are larger in the network of researchers network than in the labs network. This might indicate that blockmodeling analysis might more appropriate for the network of researchers as there is more "grouping" in this network. Out-degree centralization and betweenness centralization are larger in the network of labs. The high out-degree centralization is the result of two labs reporting many more ties than other labs. Based on this, we cannot expect a similar structure in both networks and especially not the same blockmodels and equivalences, yet we cannot rule out some similarities in structure such as similar partitions, same equivalences with different blockmodels (image matrices) etc.

Lazega et al. (2008) reported several variables measured on researchers and labs, however only specialties of researchers and labs (5 binary variables for researchers and 5 for labs) are used here for validation purposes.

|  | Res | Labs |
|---|---|---|
| Size | 98 | 78 |
| Density | 0.059 | 0.039 |
| Average in-degree | 5.745 | 3.013 |
| Centralization – degree | 0.139 | 0.220 |
| Centralization – in-degree | 0.117 | 0.118 |
| Centralization – out-degree | 0.190 | 0.381 |
| Centralization – betweenness | 0.122 | 0.244 |
| Clustering coefficient | 0.266 | 0.184 |
| Reciprocity | 0.367 | 0.083 |

**Table 1: Basic network statistics**

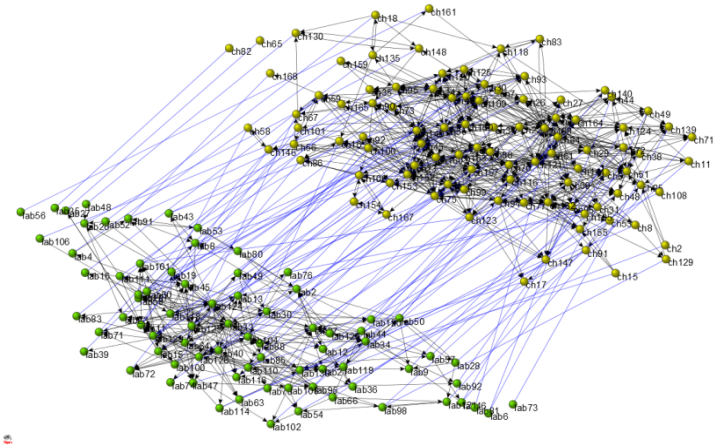

**Figure 1: Graphic representation of the whole (multilevel) network – researchers are up/right, labs are down/left**

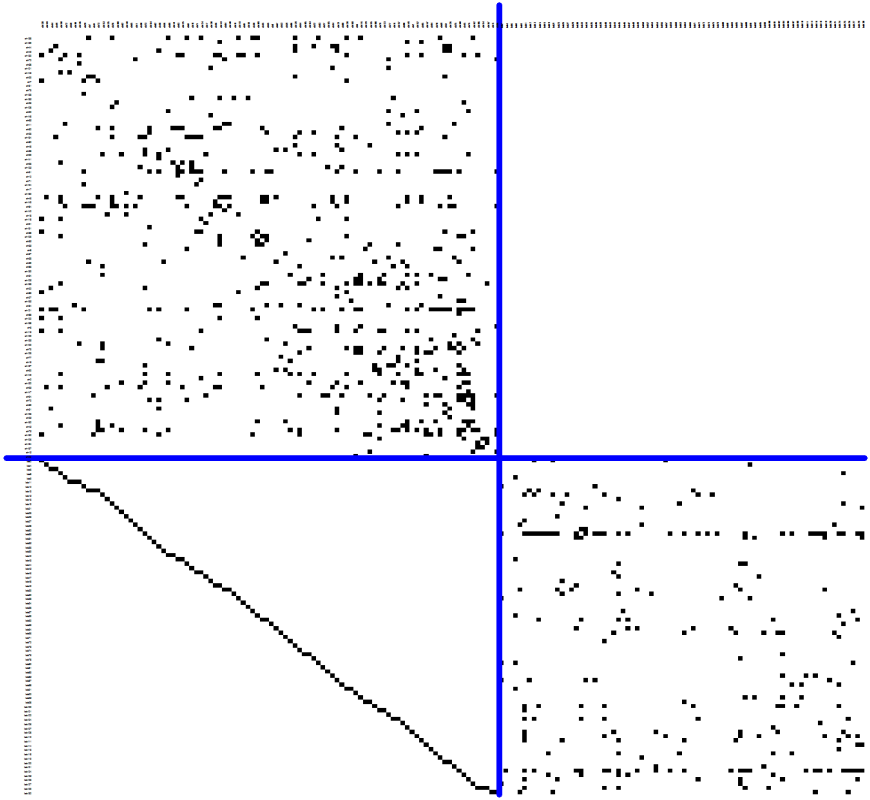

Figure 2: Matrix representation of the whole (multilevel) network – researchers are up/left, labs are down/right

To estimate the overlap of the network of researcher and network of labs networks the network of labs was reshaped to fit the network of researchers. This reshaped network of labs is actually a network among researchers where a tie between two researchers means that their labs are tied. The overall overlap measured as the percentage of researchers' ties that have "support" in the network of labs is 29.2%. If we take the opposite direction and reshape the network of researchers to labs by creating a tie between two labs if at least some researchers from those labs are connected and compute the overlap as the percentage of labs' ties that have "support" in the network of researchers we obtain 18.1%. However, here we are focusing on the first case where we are mainly interested in the support for the researchers' ties in the network of labs. Another way to assess the tie similarity of the networks is through the association coefficient Cramer's V, which is 0.216 for the network of researchers and the reshaped network of labs. The small overall overlap and small association coefficient indicate that the networks are quite different. While it is possible that some common structure is present in both networks, it is not very likely. Especially the image matrices are expected to be very different.

## 5.2 Separate analysis

The first and simplest way to analyze multilevel networks is to analyze each level separately and then compare the results. While this is the simplest analysis, it can provide relatively rich results especially in terms of similarity of structure and should always be the first step of the analysis.

### 5.2.1 Network of researchers

As mentioned, cohesive groups are first searched for in the network of researchers. The pre-specified value was set to twice the density, namely to 0.12.

As the appropriate number of clusters is not known, the number of clusters from 2 to 8 was tested and the corresponding errors are presented in Figure 3. Networks/matrices partitioned according to solutions with 4 to 7 clusters are presented in Figure 4. I excluded partitions with less or more clusters based on the desired level of complexity and results in Figure 3. Based on Figure 3 and Figure 4, the most appropriate number of clusters is 4, 5 or 7 clusters. I opted to present the 4-cluster solution as the least complex one. The same procedure for determining the appropriate number of clusters was used in the analysis of the networks of labs and in the conversion approach (presented in Section 5.3), although there the figures similar to Figure 3 and Figure 4 are omitted and only the network partition according to the selected number of clusters is presented.

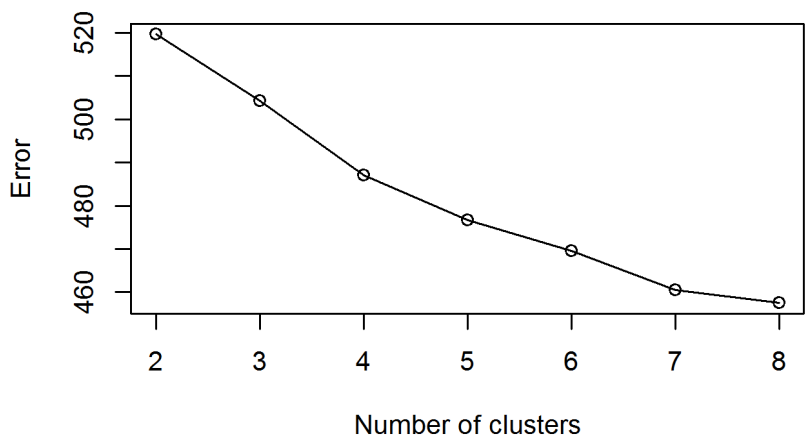

**Figure 3: Errors for SS blockmodeling of the network of researchers using a cohesive groups pre-specified blockmodel by different numbers of clusters**

The image in Figure 5 represents the densities of the resulting blocks. We can see that we have two more "cohesive" clusters (1 and 3) and two less "cohesive" ones (2 and 4). In Table 2 we explore if this partition can be associated with exogenous variables. We can see that the more cohesive clusters according to the blockmodel are also more homogeneous according to the researchers' specialties, as 91% of researchers from cluster 1 list "fundamental research" among their specialties and 92% of researchers from cluster 3 list "hematology" among their specialties. Yet this is not always the case as e.g. in the 7-cluster partition some clusters are relatively homogeneous according to specialties (not presented here) and not according to the blockmodel and vice versa.

### 5.2.2 Network of labs

The same procedure applied to the network of researchers was also applied to the network of labs. The pre-specified value was again chosen to be twice the density, that is 0.08. The number of clusters from 2 to 8 was tested and the 3-cluster solution was selected. The partitioned network and corresponding image are presented in Figure 6. The densities in the image show that the cohesive groups' model does not fit and a more core-periphery-like structure emerges. However, enforcing a core-periphery structure does not produce satisfactory results. As this example is only used for illustration I do not extend it further. In Table 3 we explore if this partition can be associated by exogenous variables, but no clear association can be found, although some differences among clusters do exist.

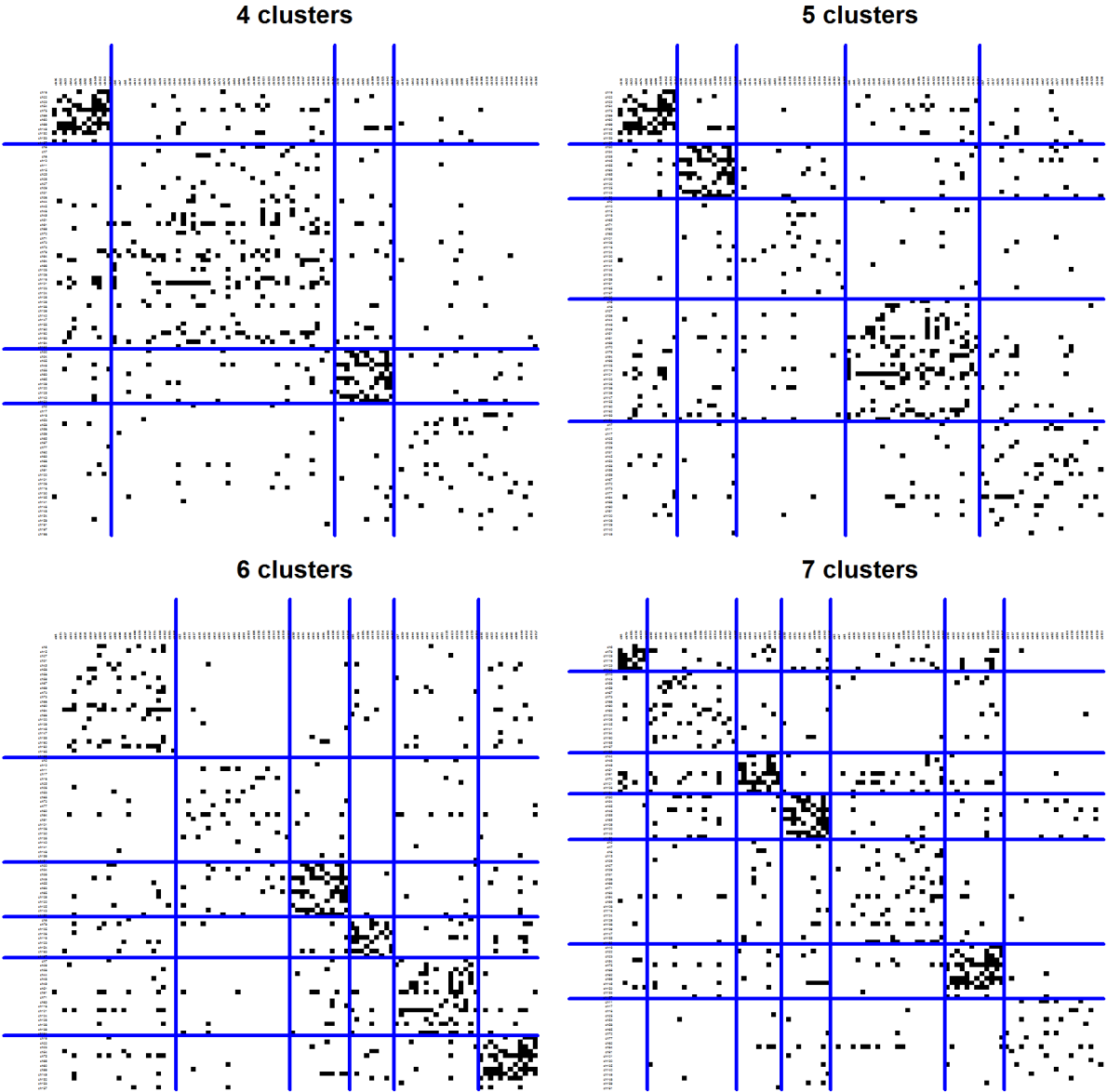

**Figure 4: The network of researchers partitioned using SS blockmodeling with a cohesive groups pre-specified blockmodel**

### 5.2.3 Comparison

Here the partitions obtained on both levels are compared. To facilitate the comparison, the labs' partition is first expanded to researchers (each researcher is "assigned" the cluster of their lab). The 2 to 8 cluster labs' partitions were compared to the 2 to 8 cluster researchers' partitions using the Adjusted Rand Index (ARI) (Hubert and Arabie, 1985). All ARIs were close to 0, the highest being 0.20 for an 8 cluster researchers' partition and 4-cluster labs' partition. Therefore, the association there among partitions based on different levels is low. This does not give much hope with regard to more complex analyses.

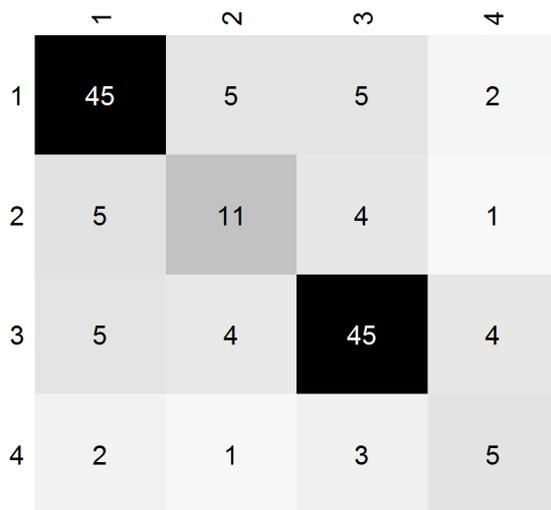

* all values in cells were multiplied by 100

Figure 5: Image of the 4-cluster partition for the network of researchers SS partition using a cohesive groups pre-specified blockmodel

|  | 1 | 2 | 3 | 4 | All |
|---|---|---|---|---|---|
| frequency | 12 | 45 | 12 | 29 | 98 |
| res - solid tumors | 0.45 | 0.56 | 0.17 | 0.38 | 0.44 |
| res - hematology | 0.18 | 0.16 | 0.92 | 0.28 | 0.29 |
| res - surgery | 0.00 | 0.18 | 0.00 | 0.00 | 0.08 |
| res - public health | 0.36 | 0.20 | 0.00 | 0.03 | 0.14 |
| res - laboratory research | 0.73 | 0.36 | 0.25 | 0.62 | 0.46 |
| res - fundamental research | 0.91 | 0.22 | 0.50 | 0.59 | 0.44 |
| lab - solid tumors | 0.55 | 0.30 | 0.18 | 0.37 | 0.33 |
| lab - hematology | 0.18 | 0.14 | 0.45 | 0.22 | 0.20 |
| lab - surgery | 0.00 | 0.07 | 0.00 | 0.00 | 0.03 |
| lab - public health | 0.00 | 0.20 | 0.00 | 0.07 | 0.12 |
| lab - laboratory research | 0.36 | 0.36 | 0.36 | 0.59 | 0.43 |
| lab - fundamental research | 1.00 | 0.45 | 0.45 | 0.67 | 0.58 |

Table 2: Averages of exogenous variables by blocks for the network of researchers SS partition using a cohesive groups pre-specified blockmodel

There is some similarity in terms of the association among the exogenous variables and the partitions. Both the researchers' and the labs' partition are to some extent associated with specific specialties, although for the network of researchers these are researchers' specialties (hematology and solid tumors), while for the labs these are labs' specialties (fundamental research and solid tumors). This does give some hope for the further analysis.

Another way to compare partitions among levels is to use a partition from one level and apply it to another level. For example, we could force the labs' partition onto the network of researchers and check the fit. For illustration the 3-cluster labs' partition obtained in the previous sub-subsection is forced onto the network of researchers. Like before when computing the ARI, here we also must first expand the labs' partition to the researchers. The network of researchers partitioned according to this partition and the corresponding image are shown in Figure 7. The image matrix shows that the densities of the on-diagonal blocks are larger than those of the off-diagonal blocks expected for the cohesive groups model; however, all are relatively close to the density of the whole network. The error for this model is 542.8, which is relatively close to the "maximal" error of 563 (obtained if the whole network is in a single null or complete block) and much further from the optimal result obtained in the sub-subsection 5.2.1, which is 504.3 for the 3-cluster partition. This indicates that, while there is some similarity among the structure of both networks, it is very small as this error is closer to "maximal" (and therefore also a "random" error) than to the optimal one.

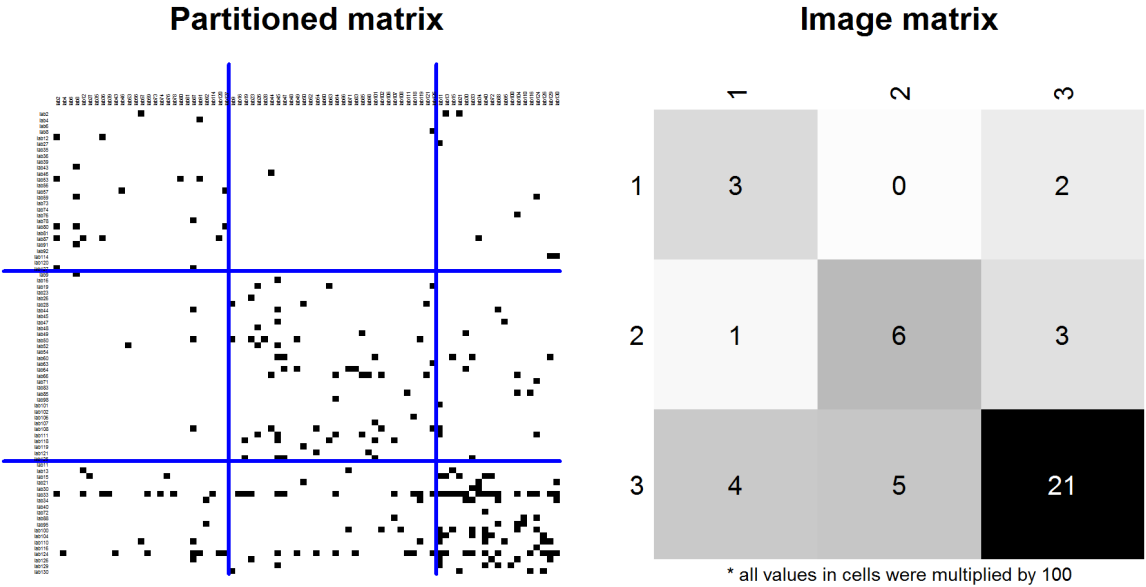

Figure 6: The network of labs partitioned into 3 clusters using SS blockmodeling with a cohesive groups pre-specified blockmodel and the corresponding image

|  | 1 | 2 | 3 | All |
|---|---|---|---|---|
| frequency | 27 | 32 | 19 | 78 |
| res - solid tumors | 0.44 | 0.58 | 0.32 | 0.47 |
| res – hematology | 0.26 | 0.27 | 0.31 | 0.27 |
| res – surgery | 0.07 | 0.17 | 0.00 | 0.10 |
| res - public health | 0.04 | 0.16 | 0.32 | 0.15 |
| res - laboratory research | 0.48 | 0.41 | 0.56 | 0.47 |
| res - fundamental research | 0.44 | 0.34 | 0.61 | 0.44 |
| lab - solid tumors | 0.22 | 0.53 | 0.22 | 0.35 |
| lab – hematology | 0.15 | 0.22 | 0.28 | 0.21 |
| lab – surgery | 0.04 | 0.06 | 0.00 | 0.04 |
| lab - public health | 0.19 | 0.03 | 0.17 | 0.12 |
| lab - laboratory research | 0.52 | 0.34 | 0.44 | 0.43 |
| lab - fundamental research | 0.52 | 0.50 | 0.78 | 0.57 |

Table 3: Averages of exogenous variables by clusters for the network of labs SS partition using a cohesive groups pre-specified blockmodel. Averages are computed as averages of average lab values among the interviewed researchers.

A similar analysis could also be performed for other partitions. In the case of applying a researchers' partition to the network of labs, reshaping this partition is a little more problematic although several approaches are reasonable. Another option to "circumvent" this is to reshape the network of labs to the researchers which is less complicated. Further discussion of this exceeds the scope of this article.

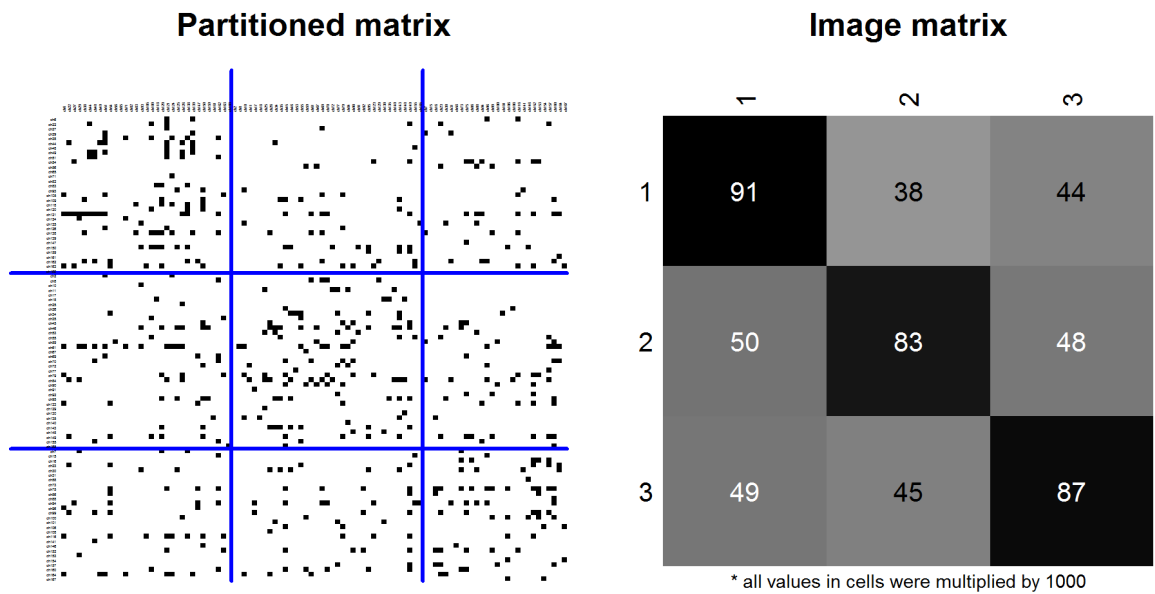

**Figure 7:** The network of researchers partitioned according to the 3-cluster labs' partition using SS blockmodeling with a cohesive groups pre-specified blockmodel and the corresponding image

## 5.3 Conversion of the multilevel problem to a classical one-level blockmodeling problem

In this subsection the multilevel problem was converted to a one-level problem, namely to a single set of units. In particular, here the network of labs was converted to researchers' "space"[14] by defining a new relation between researchers based on ties between labs. In this new relation (let us call it "institutional") two researchers are tied if their labs are tied (or if they are members of the same lab). Further analysis varies on how we combine this network with the "original" network of researchers. The first option is to create a new single-relational ("extended") network where two researchers are tied if they are tied directly ("original" network of researchers) or through their labs ("institutional" network). Such networks are also discussed by Lazega et al. (2013) in terms of extended opportunity structures. Another approach is to combine these two relations into a multi-relational network (of researchers).

### 5.3.1 Single-relational network

The same pre-specified blockmodel as was applied to the network of researchers in the previous section ("Separate analysis") was applied to this "extended" network with only the pre-specified value for constrained complete blocks (on the diagonal of the pre-specified blockmodel) being updated to 0.18, the mean of the "extended" network. A 4-cluster solution was selected as the most appropriate. The "extended" network of researchers and its "components" (the "original" network of researchers and the "institutional" network of researchers) and the corresponding images (block densities) are presented in Figure 8 (partitioned matrices on the left and image matrices on the right). In Table 4 we can see that the obtained clusters differ quite significantly, especially in the researchers' specialties.

---

[14] Conversion of the network of researchers to the labs' "space" is also possible, although more complex.

### 5.3.2 *Multi-relational network*

We will again try to search for cohesive groups in this multi-relational network (of researchers) by imposing a pre-specified cohesive groups blockmodel on both relations, where the pre-specified value is set to approximately twice the mean of each relation (0.12 for the "original" and 0.09 for the "institutional" relation). The 4-cluster solution was selected as the most appropriate. Both relations partitioned according to this solution and the corresponding images are presented in Figure 9 (partitioned matrices on the left and image matrices on the right). The densities show that the cohesive groups model fits. Moreover, we can see that clusters 1 and to a smaller extent 2 are primarily "determined" by the "original" network, while cluster 4 is chiefly defined by the "institutional" network. In Table 5 we can see that the obtained clusters differ quite significantly in specialties. In both approaches we can notice that almost all researchers in cluster 1 specialize in hematology, while most of the labs in which researchers from cluster 4 are employed specialize in fundamental research.

## 5.4 A true multilevel approach

The true multi-relational approach is an approach where we partition the multilevel network as presented in Figure 1 and Figure 2. Here a cohesive groups pre-specified blockmodel (the same as in the separate analysis stage) was used on both levels. On the two-mode network linking the two levels SS blockmodeling according to structural equivalence with constrained complete blocks was used. The complete blocks were constrained by setting the pre-specified value to 0.03 (twice the density, rounded upwards). This was used to try to give some incentive for the blocks in the two-mode network to be either (completely or almost completely) null or denser than the whole two-mode network. What we want is as many completely or nearly completely null blocks as possible to make the comparison of the researchers' and labs' clusters easier, although we do not want to force the researchers' and labs' clusters to match perfectly (e.g. by forcing all researchers from labs from a given cluster of labs to be in the same cluster of researchers). Since when using this approach finding the global (and not local) optimum is more problematic, at least 10,000 random starting points were used (instead of the 1,000 used in the other examples).

For a true multilevel approach, we have to somehow allow for an appropriate contribution of both levels and of the two-mode network. In the suggested approach, this is achieved through appropriate weighting. I decided to weight the relations (that is both levels and the two-mode network) reversely proportional to the "worst case" error, that is the error obtained in the case of only one cluster (using the blockmodeling approach selected for a given relation/level). Therefore, the following weights were used: 1 for the network of researchers, 2.346 for the network of labs and 5.478 for the two-mode network ("the original"). In order to try to obtain even clearer associations among the researchers' clusters and labs' clusters, weights with a double weight for the two-mode network was also tried ("double two-mode").

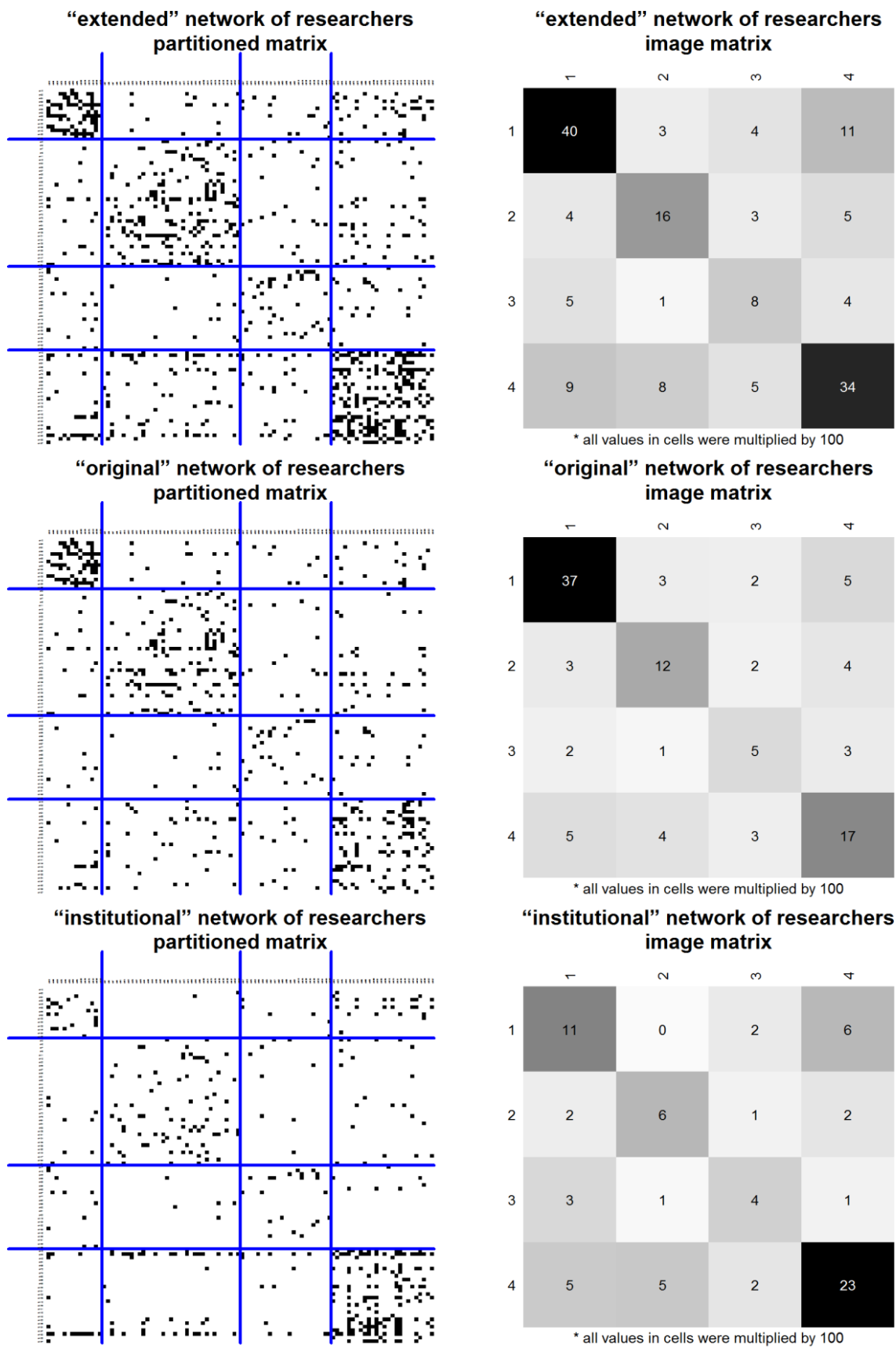

**Figure 8: The "extended" network of researchers and its "components" (the "original" network of researchers and the "institutional" network of researchers) – partitioned into 4 clusters using SS blockmodeling using a cohesive groups pre-specified blockmodel and the corresponding images**

|  | 1 | 2 | 3 | 4 | all |
|---|---|---|---|---|---|
| frequency | 14 | 35 | 23 | 26 | 98 |
| res - solid tumors | 0.21 | 0.66 | 0.39 | 0.32 | 0.44 |
| res – hematology | 0.86 | 0.11 | 0.30 | 0.20 | 0.29 |
| res – surgery | 0.00 | 0.17 | 0.09 | 0.00 | 0.08 |
| res - public health | 0.07 | 0.17 | 0.13 | 0.16 | 0.14 |
| res - laboratory research | 0.29 | 0.26 | 0.57 | 0.76 | 0.46 |
| res - fundamental research | 0.50 | 0.17 | 0.43 | 0.80 | 0.44 |
| lab - solid tumors | 0.15 | 0.35 | 0.52 | 0.22 | 0.33 |
| lab – hematology | 0.54 | 0.09 | 0.22 | 0.17 | 0.20 |
| lab – surgery | 0.00 | 0.06 | 0.04 | 0.00 | 0.03 |
| lab - public health | 0.00 | 0.26 | 0.09 | 0.00 | 0.12 |
| lab - laboratory research | 0.31 | 0.32 | 0.39 | 0.70 | 0.43 |
| lab - fundamental research | 0.46 | 0.44 | 0.57 | 0.87 | 0.58 |

**Table 4:** Averages of exogenous variables by blocks for the "extended" network of researchers SS partition using a cohesive groups pre-specified blockmodel

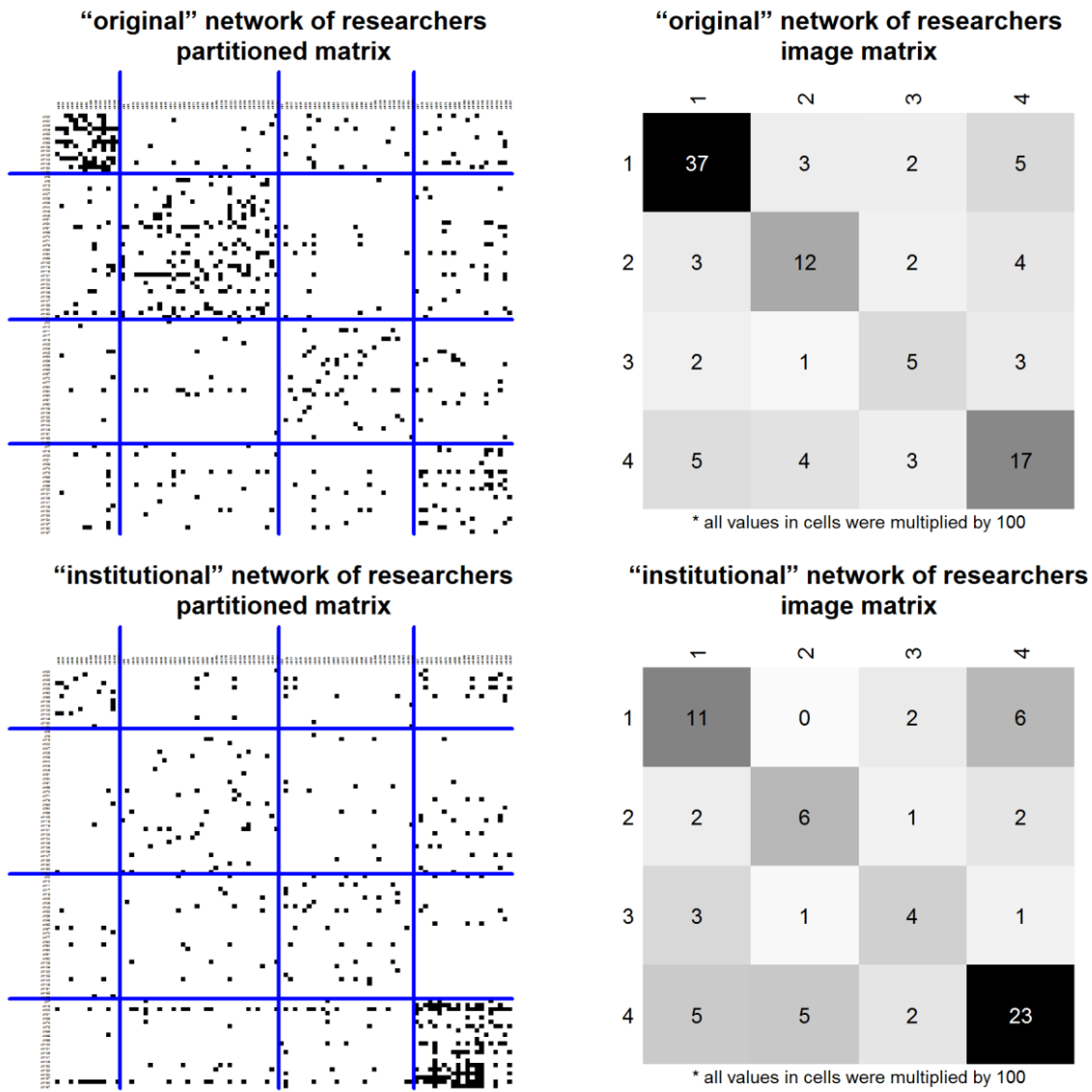

**Figure 9:** The multi-relational network of researchers partitioned using SS blockmodeling using a cohesive groups pre-specified blockmodel and the corresponding images

|                          | 1    | 2    | 3    | 4    | all  |
|--------------------------|------|------|------|------|------|
| frequency                | 14   | 34   | 29   | 21   | 98   |
| res - solid tumors       | 0.14 | 0.53 | 0.55 | 0.35 | 0.44 |
| res – hematology         | 0.93 | 0.12 | 0.21 | 0.25 | 0.29 |
| res – surgery            | 0.00 | 0.18 | 0.07 | 0.00 | 0.08 |
| res - public health      | 0.00 | 0.09 | 0.24 | 0.2  | 0.14 |
| res - laboratory research| 0.21 | 0.38 | 0.48 | 0.75 | 0.46 |
| res - fundamental research| 0.50| 0.26 | 0.34 | 0.85 | 0.44 |
| lab - solid tumors       | 0.15 | 0.24 | 0.55 | 0.28 | 0.33 |
| lab – hematology         | 0.54 | 0.00 | 0.21 | 0.33 | 0.20 |
| lab – surgery            | 0.00 | 0.06 | 0.03 | 0.00 | 0.03 |
| lab - public health      | 0.00 | 0.27 | 0.07 | 0.00 | 0.12 |
| lab - laboratory research| 0.31 | 0.52 | 0.34 | 0.50 | 0.43 |
| lab - fundamental research| 0.46| 0.52 | 0.48 | 0.94 | 0.58 |

**Table 5: Averages of exogenous variables by blocks for the multi-relational network of researchers SS partition using a cohesive groups pre-specified blockmodel**

Due to the time complexity of the algorithm, the size of the multilevel network and space limitations of this article, I fixed the number of clusters to 4 researchers' clusters and 3 labs' clusters. These two numbers were selected based on the results of the separate analysis stage. The partitioned multilevel network and corresponding images using the "original" weights and using "double two-mode" weights are presented in Figure 10 (partitioned matrices on the left and image matrices on the right).

As expected, the two-mode network is better partitioned (fewer "in-between" blocks) when "double two-mode" weighting is used ("double" weight is given to the two-mode network). However, due to this additional emphasis on a clearer two-mode network, the diagonal (complete) blocks in the network of researchers have lower densities. Only the partition obtained with "double two-mode" weighting will be further inspected. In fact, the results indicate that maybe some "in-between" weighting would be desired[15] or that "double two-mode" weighting should be used to further optimize the "original" weighting solution. However, as this article's emphasis is not on results, we do not explore these options further.

We can notice that researchers from cluster 1 are mostly in labs from cluster 7, all of those from cluster 2 are in labs from clusters 7, and so on. Similarly, the labs from cluster 7 mainly employ researchers from clusters 1 and 2. The correspondence among the researchers' and labs' clusters is not one-to-one, yet it is clear that units of a certain level that are in the same cluster are predominantly connected to units from another level that are in one or two clusters.

The association among both the (researchers' and labs') partitions and exogenous variables is examined in Table 6. We can notice that many clusters have a large share of researchers or labs with certain specialties. We first look at the researchers' clusters. Here 93% of researchers in cluster 1 specialize in hematology and 65% of researchers in cluster 2 specialize in solid tumors. 95% of labs associated with researchers from cluster 4 do fundamental research, and so do 82% of these researchers. Only cluster 3 is not dominated by a certain specialty (although it has an above-average number of researchers specializing in solid tumors and surgery).

---

[15] It would be even better if the multiobjective approach suggested by Brusco et al. (2013) were used.

The concentration in the labs' clusters is not as high. The highest concentration can be found in cluster 5 where 93% of labs and 83% of researchers[16] do fundamental research. In cluster 7, 47% of researchers and 50% of labs specialize in solid tumors, while 39% of researchers and 31% of labs specialize in hematology. Cluster 6 contains an above-average share of labs that specialize in surgery and public health. We can notice that more "concentrated" clusters have higher densities in corresponding diagonal blocks. These characteristics of labs' clusters are expected if we observe to which researchers' clusters these labs' clusters are connected.

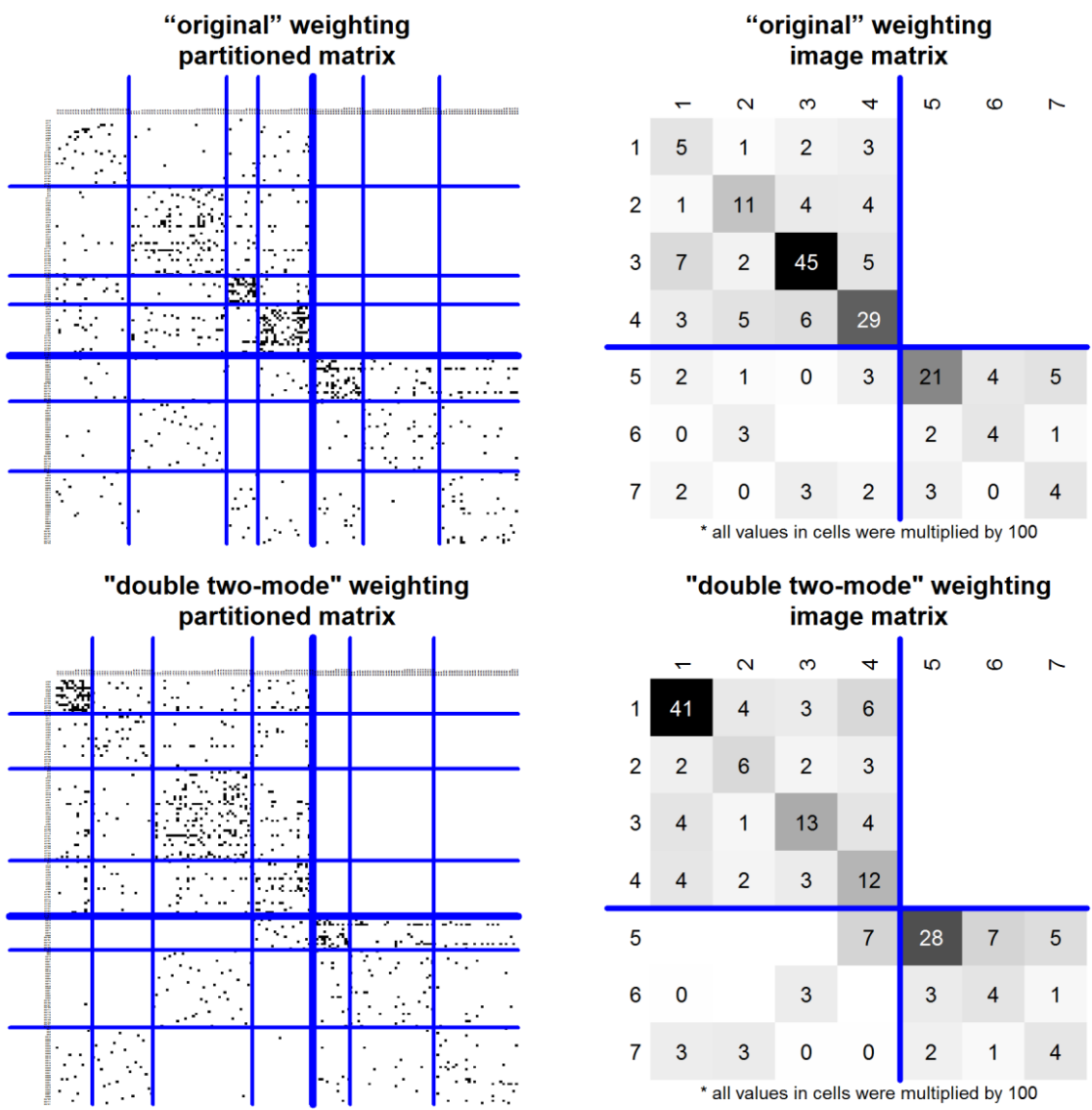

Figure 10: Multilevel network partitioned using SS blockmodeling using a cohesive groups pre-specified blockmodel. Weighting is indicated above the matrices.

---

[16] In fact, the average share of researchers interviewed within these labs with this specialty is 83%.

|  | Researchers | | | | | Labs | | | |
|---|---|---|---|---|---|---|---|---|---|
|  | 1 | 2 | 3 | 4 | all | 5 | 6 | 7 | all |
| Freq | 14 | 23 | 38 | 23 | 98 | 14 | 32 | 32 | 78 |
| res - solid tumors | 0.14 | 0.65 | 0.50 | 0.32 | 0.44 | 0.37 | 0.52 | 0.47 | 0.47 |
| res - hematology | 0.93 | 0.17 | 0.13 | 0.27 | 0.29 | 0.35 | 0.12 | 0.39 | 0.27 |
| res - surgery | 0.00 | 0.09 | 0.16 | 0.00 | 0.08 | 0.00 | 0.16 | 0.08 | 0.10 |
| res - public health | 0.00 | 0.22 | 0.11 | 0.23 | 0.14 | 0.29 | 0.09 | 0.16 | 0.15 |
| res - laboratory research | 0.21 | 0.52 | 0.37 | 0.73 | 0.46 | 0.69 | 0.36 | 0.48 | 0.47 |
| res - fundamental research | 0.50 | 0.39 | 0.24 | 0.82 | 0.44 | 0.83 | 0.30 | 0.42 | 0.44 |
| lab - solid tumors | 0.15 | 0.65 | 0.22 | 0.30 | 0.33 | 0.29 | 0.23 | 0.50 | 0.35 |
| lab - hematology | 0.54 | 0.22 | 0.03 | 0.30 | 0.20 | 0.29 | 0.06 | 0.31 | 0.21 |
| lab - surgery | 0.00 | 0.04 | 0.05 | 0.00 | 0.03 | 0.00 | 0.06 | 0.03 | 0.04 |
| lab - public health | 0.00 | 0.00 | 0.27 | 0.05 | 0.12 | 0.07 | 0.26 | 0.00 | 0.12 |
| lab - laboratory research | 0.31 | 0.30 | 0.54 | 0.45 | 0.43 | 0.50 | 0.52 | 0.31 | 0.43 |
| lab - fundamental research | 0.46 | 0.48 | 0.49 | 0.95 | 0.58 | 0.93 | 0.48 | 0.50 | 0.57 |

Table 6: Averages of exogenous variables by clusters for the multilevel network SS partition using a cohesive groups pre-specified blockmodel with "double two-mode" weighting. Averages for the labs' clusters are computed as averages of the average lab values among the interviewed researchers.

When comparing the results we could say that researchers' clusters 1 and 2 are mainly influenced by researchers' characteristics, while clusters 3 and 4 are largely determined by the lab clusters to which they are tied.

## 5.5 Comparison of the results using different approaches

In this section, several approaches were used on the two-level network of cooperation among researchers and labs. Although different approaches are not designed to produce the same results, some results from different approaches are compared in this section. Of course, not every possible comparison is presented here.

One of the results that is common to all approaches is the partition of a researchers into cohesive groups and corresponding blockmodels of the network of researchers. In the separate analysis approach, this partition is found by only taking the ties among the researchers into account. In the other approaches, the ties among laboratories and the membership of researchers in laboratories are also taken into account (see the previous section for exactly how they are accounted for). When using the separate analysis approach and both versions of the conversion approach, the suitable number of clusters was estimated by looking at how the inconsistency of the model decreases when the number of clusters increases. In all these cases, 4 clusters were selected as the most appropriate. As a consequence, 4 clusters of researchers were also used in the true multilevel approach. In addition to the 4-cluster partitions of researchers, 3-cluster partitions of labs were also obtained in the separate analysis (based on the network of labs only) and in the true multilevel approach (by also taking the network of researchers and the two-mode network into account). These labs' partitions were expanded to researchers (each researcher is "assigned" to the cluster of their lab).

In Table 7 the similarities between all of these partitions (the 4-cluster partitions of researchers and the 3-cluster partitions of labs)[17] is measured by ARI. None of the partitions are essentially the same, although practically all indicate similarity above that expected by chance. While most of these values would be considered low by Steinley (2004), these indices are not used here to measure recovery of the "true" cluster structure as used by Steinley (2004), but just the similarity of the partitions.

The most similar pair of 4-cluster partitions of researchers is composed of partitions returned by the multi-relational conversion approach and that obtained by the true multilevel approach with "double two-mode" weighting (ARI = 0.77, moderate recovery according to Steinley (2004)). These similarities are expected (as mentioned in subsection 4.3.1) since the "double two-mode" weighting results in most lower lever clusters being tied to only one higher level cluster and vice versa (as much as possible due to the different number of lower and higher level clusters).

|  | Res | Labs | S-con | M-con | Res-ML | Labs-ML | Res-ML2 | Labs-ML2 |
|---|---|---|---|---|---|---|---|---|
| **Res**: Researchers – Separate analysis | 1 | 0.01 | 0.57 | 0.31 | 0.55 | 0.23 | 0.37 | 0.24 |
| **Labs**: Labs – Separate analysis | 0.01 | 1 | 0.15 | 0.35 | 0.06 | 0.45 | 0.25 | 0.30 |
| **S-con**: Single-relational conversion approach | 0.57 | 0.15 | 1 | 0.46 | 0.35 | 0.29 | 0.44 | 0.37 |
| **M-con**: Multi-relational conversion approach | 0.31 | 0.35 | 0.46 | 1 | 0.42 | 0.34 | 0.77 | 0.60 |
| **Res-ML**: True multilevel approach ("original" weighting) – researchers | 0.55 | 0.06 | 0.35 | 0.42 | 1 | 0.32 | 0.38 | 0.23 |
| **Labs-ML**: True multilevel approach ("original" weighting) – labs | 0.23 | 0.45 | 0.29 | 0.34 | 0.32 | 1 | 0.28 | 0.31 |
| **Res-ML2**: True multilevel approach ("double two-mode" weighting) – researchers | 0.37 | 0.25 | 0.44 | 0.77 | 0.38 | 0.28 | 1 | 0.76 |
| **Labs-ML2**: True multilevel approach ("double two-mode" weighting) – labs | 0.24 | 0.30 | 0.37 | 0.60 | 0.23 | 0.31 | 0.76 | 1 |

Table 7: Similarity of the 4-cluster partitions of researchers obtained with different approaches measured by ARI

The similarities in Table 7 also reveal some other properties of the methods. Most approaches that take both levels into account produce partitions that are more similar to the separate analysis of researchers' and labs' partitions than would be expected by chance. Now let us examine more closely the similarities of the true multilevel partitions (also with other partitions). As expected, the similarity of the researchers' and labs' partitions is much greater when the "double two-mode" weighting was used (as opposed to the "original" weighting) since the ties between the clusters from different levels are much higher in this case. However, due to the increased emphasis on the blockmodel of the two-mode network, there is less similarity of the researchers' and labs' partitions with the corresponding partitions from the separate analysis approach.

---

[17] It should be noted that the computed similarities can be drastically different if we select a different number of clusters. E.g., the 2-cluster partition single-relational conversion approach partition is much more similar to the 2-cluster labs' partition (ARI = 0.45) than to the 2-cluster researchers' partition (ARI = -0.01) (for the 4-cluster partitions presented in Table 7 the situation is reversed). Also in the case of 2-cluster partitions both conversion approaches produce the same partition.

In addition to partitions we can also compare obtained image matrices (or blocks in general). In all cases (see Figure 5, Figure 8, Figure 9 and Figure 10), the obtained image is compatible with the cohesive groups model. The densities of the diagonal blocks are much higher than the densities of the off-diagonal blocks, with the exception of the diagonal block with the lowest density since one or two off-diagonal blocks have a similar density in most cases. In the case of separate analysis, two (out of four) diagonal blocks are relatively dense (with densities above 0.4), one more block with a clearly above-average density and one block with about average density. For other solutions (those also taking the other level into account), the image is similar except that on the diagonal we have only one relatively dense (with densities above 0.4) block and two blocks with a clearly above-average density. Therefore, all images are relatively similar.

The characteristics of the obtained clusters in terms of the researchers' and labs' specialties (see Table 2, Table 4, Table 5 and Table 6) reveal that the most cohesive cluster[18] (i.e. having the diagonal block with the highest density) is always composed of predominantly researchers specializing in hematology. One of the blocks with above-average densities is primarily composed of researchers specializing in fundamental research (and who are employed in labs specializing in fundamental research). In the conversion approach this cluster is also the most cohesive cluster in the "institutional" network (meaning that the labs of these researchers are relatively strongly connected), while in the true multilevel approach this cluster is strongly (in the case of "double two-mode" weighting almost exclusively) connected to the most cohesive labs' cluster. A similar analysis could also be performed for the labs.

## 5.6 Lessons learned from the application

In this section several approaches to blockmodeling multilevel networks were applied to the multilevel network of elite cancer researchers in France. Yet in most cases these approaches should be seen as complementary rather than as alternatives. The separate analysis should be the first stage of any blockmodeling attempt on multilevel networks. In this application we saw that cohesive groups can be found at both levels, although only some of the groups found can truly be labeled cohesive. In most cases, the more cohesive the groups are, the more they are "concentrated" on the given researchers' or labs' specialty. However, when partitions for both levels were compared, not much overlap (association) was found. While this gave little hope for the usefulness of the multilevel approaches, the results of both analyses of the "combined" or "extended" networks (the conversion approach) and of the true multilevel approach showed that even in such cases multilevel approaches can be useful.

By using the "conversion" approach we showed how different levels can be combined into a single-level network to obtain a partition based on both (all) levels. Especially the results of the multi-relational version showed that some of the clusters obtained were determined more by one level and some by another.

---

[18] One of the two most cohesive clusters in the separate analysis case

Using the true multilevel approach gave us two partitions, one for each level. While these partitions are individually not as "optimal" as those from the separate analysis, we also obtained a partition of the two-mode network showing how they are related. For example, labs' cluster 5, where most labs do fundamental research, is composed of most researchers from cluster 4. On the other hand, lab cluster 7 is composed of researchers from researcher clusters 1 and 2. While the characteristics and researchers' ties show that they should be in different researchers' clusters, the similarity or better said sparsity of their labs' ties put them in the same lab cluster (7). This shows that the true multilevel approach might be a good compromise between a separate analysis, where there might be no relation among partitions from different levels, and a combined approach, where the partitions are functionally linked. It provides partitions somewhat tailored to individual levels but with clear linkages among clusters from different levels. In this application, a little less weight should probably be given to the two-mode network to allow the partitions to be more tailored to the individual levels, although this option is not explored further as it would exceed the scope of this article.

# 6 Conclusions

In the article several approaches to the blockmodeling of multilevel networks were presented. First, a multilevel network was defined as a network where ties between units of each level are studied together with ties between levels. The presented approaches are a separate analysis of individual levels, followed by a comparison of results, conversion of the multilevel network to a one-level network, and the true multilevel approach where all levels and ties among them are modeled simultaneously. The article uses generalized blockmodeling as its framework, although at least the first two approaches can be implemented using any blockmodeling approach. Some extensions to generalized blockmodeling are also suggested that facilitate the use of this framework for blockmodeling multilevel networks. These extensions are, however, also useful for blockmodeling one-level networks.

The advantages and limitations of each of these approaches are discussed. While this is not the main purpose of this article and will be subject to further research, some suggestions are made regarding which approach should be used in a given situation. I suggest that a separate analysis should be used as the first stage in any blockmodeling analysis of multilevel networks. The conversions approach is most suitable when we want to focus on a certain level, while using information from the other level(s) to improve the partition and/or the other level(s) can be seen as indirect relations for units of the level in focus. In contrast, the multilevel approach should be used when we already have some knowledge about the structure of the network. One benefit of using this approach is that it can provide us with a novel insight into ties among clusters from different levels. It can also help us search for such clusters at individual levels where the ties among them are relatively "clean". In addition, using the multilevel approach can have similar effects as the conversion approach since information from one level is used to better determine clusters on the other level.

To sum up, the suggested approaches enable a true multilevel blockmodeling analysis of multilevel networks.

## Acknowledgements

I would like to thank Emmanuel Lazega (and his collaborators) for introducing me to this problem, for providing me with the multilevel network to test my ideas and for discussing these ideas with me. It would not have been possible to prepare this article without his help.

## Appendix A: Possible restrictions for blockmodeling two-mode networks

While it would probably be most desirable to impose relatively vague restrictions in terms of the pattern ties in the image of the two-mode networks, like that there should be one or two ties from each row cluster and at least one tie going into each column cluster[19], that is currently not possible within generalized blockmodeling and therefore represents a possible extension.

Yet it is possible to specify possible block types for each pair of row and column clusters, which means we can say that for a given pair there must or may not be a tie and, if a tie is allowed, which kinds of ties are allowed. For example, we can specify that there must be a certain type of tie from row cluster 1 to column cluster 2 and that there may be a tie of one of three types from row cluster 1 to row cluster 3, but we cannot say that there must be one or two ties from row cluster 1 to any of the column clusters. The "vague" restrictions mentioned earlier can in principle be imposed by running the procedure several times with all pre-specifications that match those restrictions and then selecting the best result. However, this exceeds the scope of this article.

At this point, only one specific configuration of restrictions will be presented, namely the one where each row cluster must be connected to exactly one column cluster (referred to later as a "1 to 1 restriction"), which essentially means that level-one units joined in a cluster must all be affiliated to level-two units that also are all only in one cluster. As mentioned, we must exactly specify to which column cluster each row cluster must be tied, although this is not a limitation if blockmodels for one-mode networks are specified only in terms of equivalences or allowed block types and not their positions[20]. Obviously, such a restriction is only possible if the number of the row (first level) and column (second level) clusters is the same. This restriction is specified by using a pre-specified blockmodel on the two-mode clusters where only non-null blocks are allowed on the diagonal and only null blocks off-diagonal.

This restriction ("1 to 1 restriction") is very similar to the restrictions implicitly imposed in the conversion approach as discussed in subsection 4.2. The main difference with regard to the multi-relational conversion approach (in the case of partition type two-mode networks) lies in the fact that here we can explicitly decide how much "weight" we want to give to this restriction. In the conversion approach, this restriction is a consequence of the fact that only a partition at one level is obtained directly, while the other is obtained (if desired) by reshaping the directly obtained partition (the partition of the "other" level is a function of the "first" partition and the two-mode network)[21]. The true multilevel approach also ensures that each unit (regardless of its level) is always classified in just one cluster.

---

[19] This would mean that level-one units that are affiliated to level-two units from a given level-two cluster should be assigned to one or at most two level-one clusters.

[20] However, it is true that even if blockmodels for one-mode networks are specified only in terms of equivalences or allowed block types, optimization is more complex, meaning that more repetitions of the local search algorithm (with random starting points) are required to obtain results of the same quality.

---

[21] For the conversion approach, the "1 to 1 restriction" is total (no inconsistencies are possible) where the "higher" level is used as a base. In case the "lower" level is taken as a base in the conversion approach, the restriction is violated only if two lower level units tied to the same "higher" level units are in different clusters. The fact that all "lower" level units tied to the same "higher" level unit have an identical pattern of ties in the reshaped relation ($R^{2*}$) decreases the possibility of such cases.